\documentclass[11pt,twoside]{article}
\usepackage{asp2004}
\usepackage{psfig}
\usepackage{epsf}
\usepackage{graphics}
\usepackage{lscape}

\newcommand {\Msun}{\ensuremath{M_{\odot}}}

\newcommand {\kms}{\ensuremath{{\rm km\,s}^{-1}}}

\begin{document}

\title{Run--away formation of intermediate-mass black holes in dense star clusters}

\author{Marc Freitag}
\affil{Astronomisches Rechen-Institut, Heidelberg, Germany}
\author{M. Atakan G\"urkan and Frederic A. Rasio}
\affil{Department of Physics and Astronomy, Northwestern University, USA}

  


\section{Fast core collapse of a stellar cluster}

We are exploring pathways through which dynamical evolution of a
stellar cluster may lead to the formation of an intermediate-mass black hole
(IMBH).

\begin{figure}[!ht]
\plotone{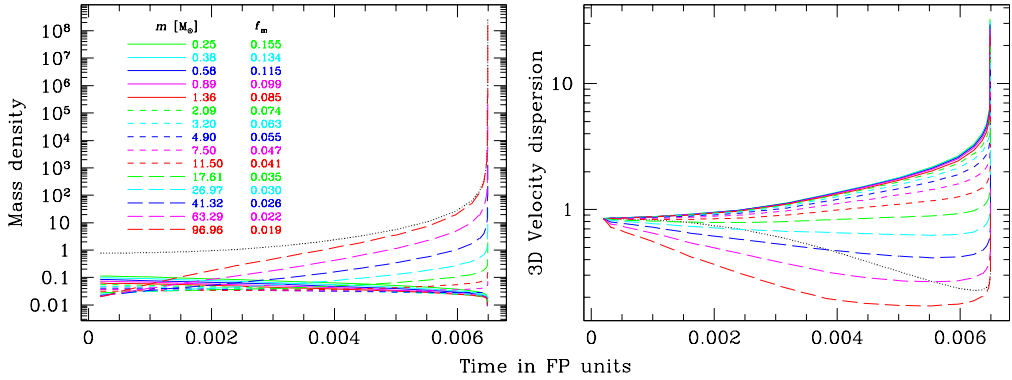}
\caption{  Core-collapse evolution of a multi-mass
  stellar cluster simulated with the gaseous model \citep{ST95}.  {Left}:
  Evolution of the central density for the 15 individual mass
  components ($m$ is the mass of the stars in the
  component, $f_m$ the fraction of the total mass). The dotted 
  line shows the total density. {Right}: Evolution of the central
  velocity dispersions. The dotted black line shows the mass-averaged
  value. $N$-body units are used for the $y$-axes. \label{fig:cc}}
\end{figure}

We consider the evolution of spherical clusters with a broad mass
function ($M_\ast=0.2-120\,\Msun$, typically). Using so-called ``Monte
Carlo'' (MC) and ``gaseous'' simulation techniques, we have shown that
core collapse, driven by mass-segregation, occurs very quickly, i.e.
within of order 15\,\% of the central relaxation time \citep*{GFR04}.
During core collapse, the central regions of the cluster become
completely dominated by the most massive stars. The central density
steadily increases until the first stellar collisions occur.  However,
the central velocity dispersion {\em decreases} during most of the
evolution as a result of a tendency toward kinetic energy
equipartition between massive stars and lighter ones, see
Fig.~\ref{fig:cc}. Hence relative velocities of a few 1000\,{\kms} are
never reached and one may expect disruptive collisions to be rare.

\section{Collisional run-away}

\begin{figure}[!ht]
\plotone{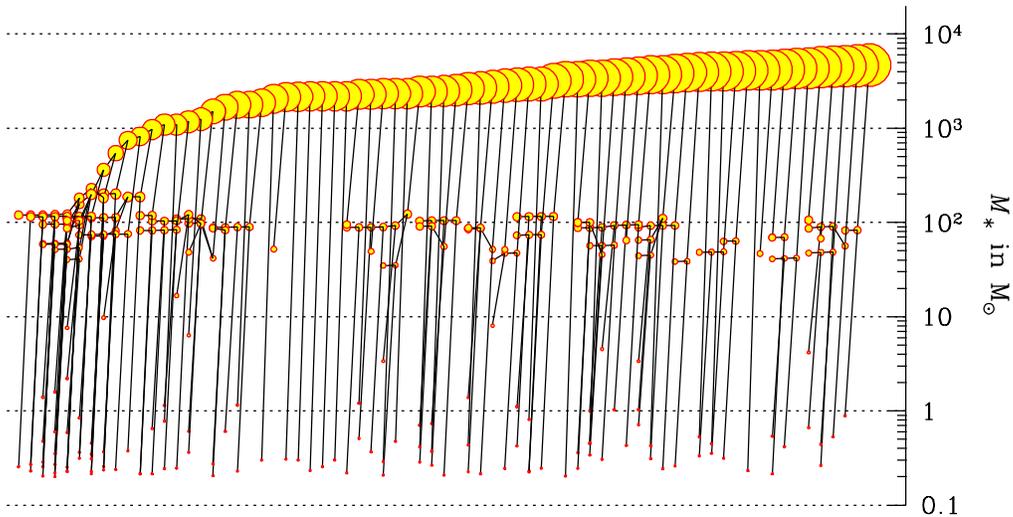}
\caption{Collisional run-away in a typical
  MC simulation with $10^6$ particles. The initial 1D velocity
  dispersion is 130\,{\kms}. Shown is the merger tree for the run-away
  VMS. Evolution is from left to right. The vertical axis indicates
  the stellar mass. Note that the largest fraction of the final
  mass of the VMS comes from stars near the top end of the IMF, around
  100\,{\Msun}. \label{fig:runaway}}
\end{figure}

In recent MC simulations, we have introduced collisions between single
MS stars (\citealt*{RFG03}; Freitag, G\"urkan \& Rasio, in
preparation). Our prescription for the outcome of collisions is based
on $\sim$\,15\,000 SPH simulations \citep{FB04}. Hence, we do
not assume that collisions are perfect mergers but allow for
collisional mass loss and fly-bys, a likely outcome for encounters
with relative velocities of a few 100\,{\kms}. However, we observe
that, provided core collapse occurs within less than $\sim$\,3\,Myrs
(the time needed for massive stars to evolve off the MS), the cluster
always enters a run-away phase in which a star more massive than
1000\,{\Msun} grows through repeated mergers (mostly with
$\sim$\,100\,{\Msun} stars). This is shown, for one simulation, in
Fig.~\ref{fig:runaway}. Such a very massive star (VMS) is a likely
progenitor for an IMBH.

\acknowledgments{We thank Rainer Spurzem for his help with the gase\-ous
  model. The work of MF is supported by the scheme SFB-439/A5 of the
  German Science Foundation (DFG). The work of AG and FR is supported
  by NASA ATP Grant NAG5-12044 and NSF Grant AST-0206276.}


\end{document}